# Two Distinct Charge Orders in Infinite-layer PrNiO$_{2+\delta}$ revealed by Resonant X-ray Diffraction

Xiaolin Ren[1,2], Ronny Sutarto[3], Qiang Gao[1], Qisi Wang[4], Jiarui Li[5], Yao Wang[6], Tao Xiang[1,2,7], Jiangping Hu[1,2], J. Chang[4], Riccardo Comin[5], X. J. Zhou[1,2,7,8,*], and Zhihai Zhu[1,2,8,*]

[1]*Beijing National Laboratory for Condensed Matter Physics, Institute of Physics, Chinese Academy of Sciences, Beijing 100190, China*

[2]*School of Physical Sciences, University of Chinese Academy of Sciences, Beijing 100049, China*

[3]*Canadian Light Source, Saskatoon, Saskatchewan S7N 2V3, Canada*

[4]*Physik-Institut, Universität Zürich, Winterthurerstrasse 190, CH-8057 Zürich, Switzerland*

[5]*Department of Physics, Massachusetts Institute of Technology, Cambridge, Massachusetts 02139, USA*

[6]*Department of Physics and Astronomy, Clemson University, Clemson, SC 29631, USA*

[7]*Beijing Academy of Quantum Information Sciences, Beijing 100193, China*

[8]*Songshan Lake Materials Laboratory, Dongguan 523808, China*

[*]*To whom correspondence should be addressed.*

*Emails: XJZhou@iphy.ac.cn, zzh@iphy.ac.cn*

(Dated: November 24, 2023)

**A broken translation symmetry has recently been revealed in infinite-layer nickelates, which has piqued considerable interest in its origin and relation to superconductivity, as well as in its comparison to charge order in cuprates. Here, by performing resonant x-ray scattering measurements in thin films of infinite-layer $PrNiO_{2+\delta}$, we find that the superlattice reflection at the Ni $L_3$ absorption edge differs considerably from that at the Pr $M_5$ resonance in their dependence on energy, temperature, and local symmetry, indicating they are two distinct charge orders despite the same in-plane wavevectors. These dissimilarities might be related to the excess oxygen dopants, considering that the resonant reflections were observed in an incompletely reduced $PrNiO_{2+\delta}$ film. In addition, our azimuthal analysis suggests that the oxygen ligands should play a pivotal role in the charge modulation revealed at the Ni $L_3$ resonance.**

The realization of superconductivity in infinite-layer nickelates has finally given important answers to the long-lasting debate over whether these layered nickelates are intriguingly analogous to the high-$T_c$ cuprates [1-6], though the debate remains unsettled given that many important differences between these two material systems have recently been revealed [7-12]. A defining characteristic of cuprate superconductors is that multiple electronically ordered phases of nearly equal energy emerge by hole-doping of $CuO_2$ planes, and they usually compete or coexist with one another; for example, charge order has been revealed among various families of cuprates as a dominant competitor of superconductivity [13-22]. Recently, a broken translation symmetry has been reported for infinite-layer nickelates exhibiting a wavevector $\vec{Q} = (\sim\frac{1}{3}, 0)$ along the Ni-O bond direction in the $NiO_2$ planes [23-25]. Besides, the measurements of the scattering signal versus doping level uncover a competitive interplay between the superlattice reflection and superconductivity in $La_{1-x}Sr_xNiO_2$ [23]. These observations appear to be reminiscent of charge orders in underdoped cuprates [13-22], hinting at a possible common thread in understanding charge order phenomenology in both infinite-layer nickelates and cuprates.

However, there are essential distinctions between the superlattice reflections in infinite-layer nickelates and charge orders in underdoped cuprates. For example, the resonant reflections were reported for the nominally 'parent' phase of infinite-layer nickelates [23-25], namely at zero doping if ignoring the self-doping effect [9, 11, 26, 27], while for cuprates charge orders in the $CuO_2$ planes occur only in doped ones [13-22]. Another peculiarity of the superlattice peaks in infinite-layer nickelates at zero doping is that their onset temperatures are much higher than those of charge orders in underdoped cuprates and remain finite at room temperature [23-25, 28]. Besides, an additional complexity in infinite-layer nickelates stems from the thin film synthesis; for instance, capping layers were claimed to play an important role in the emergence of the superlattice peaks [24]. Therefore, the root of the superlattice reflections in infinite-layer nickelates remains elusive [23-25, 29-32].

Here we used resonant x-ray scattering measurements to systematically investigate the resonant reflections in $PrNiO_2$ as a function of photon energy, temperature, and azimuthal angle. We find significant dissimilarities between the resonant reflections taken at the Ni $L_3$ and the Pr $M_5$ absorption edges, which suggests fundamentally different interpretations for their origins at the two resonances. The azimuthal angle analysis reveals a predominant *d*-wave character of the resonant reflection at the Ni $L_3$ absorption edge, which somewhat resembles that in certain underdoped cuprates where charge modulates on the O-2*p* orbitals forming a bond order [33-36].

Figure 1**a** shows a schematic of the scattering geometry of our RXS measurements. Throughout this paper, we use reciprocal lattice units (r. l. u.) for the tetragonal crystal structure of $PrNiO_2$ whose axes are depicted

in Fig. 1**a**. The reciprocal space (*H*, *K*, *L*) components are given in units of ($\frac{2\pi}{a}$, $\frac{2\pi}{b}$, $\frac{2\pi}{c}$), where the in-plane lattice constants $a = b = 3.905$ Å, and the out-of-plane lattice constant $c = 3.443$ Å. The film samples are aligned with the (*H*, 0, *L*) in the scattering plane; that is, the crystallographic axis *b* is perpendicular to the scattering plane. This scattering geometry corresponds to the azimuthal angle $\phi = 0$ in the definition for the azimuthal angle dependence of the RXS intensity in the following sections. We show in Fig. 1**b** the momentum scans across the charge order $\vec{Q}_{CO} = (\sim\pm\frac{1}{3}, 0, 0.365)$ at the Ni $L_3$ resonance ($E = 851.3$ eV) for a typical film of PrNiO$_2$ with the polarizations of incident photons fixed to $\sigma$ and $\pi$ channel. The scattered intensity is much stronger in the case of the $\sigma$- than the $\pi$-polarized incoming photons, similar to that observed in cuprates [13, 15, 35]. As shown in Fig. 1**c**, similar findings have been uncovered for the momentum scans performed at the Pr $M_5$ resonance ($E = 928.4$ eV).

To further unveil the origin of the ($\sim\pm\frac{1}{3}$, 0, 0.365) reflections, we include in Fig. 2 the photon energy dependence of the scattered intensity. Here the center of the scans corresponds to constant wave vector ($\frac{1}{3}$, 0, 0.365) by adjusting both $\theta$ and $2\theta$ (detector angles) for each different photon energy. Strong energy dependence of the scattering intensity can be seen in both cases. Figure 2**a**, **b** shows a series of representative momentum scans across the ($\frac{1}{3}$, 0, 0.365) reflection with the photon energies tuned to the Ni $L_3$ and the Pr $M_5$ absorption edges at $T = 20$ K. In Fig. 2**c**, we present the integrated signals of the ($\sim\frac{1}{3}$, 0, 0.365) reflection as a function of photon energies as well as the x-ray absorption spectroscopy (XAS) measured using both $\sigma$- and $\pi$-polarized incident photons. Consistent with the previous work, the XAS at the Ni $L_3$ absorption edge exhibits a double-peak feature denoted by A and A′ [24-26, 37]. Peak A′ is prominent while using $\pi$-polarized incident photons and reflects the Pr-Ni hybridization [26, 37]. The photon energy profiles show a peak maximum at $E = 851.3$ eV, approximately 0.3 (0.7) eV below peak A′ (A). The scattered intensity of the ($\sim\frac{1}{3}$, 0, 0.365) reflection exhibits strong energy dependence near the Pr $M_5$ absorption edge, and the integrated intensity of the reflection peaks at $E = 928.4$ eV, which is about 0.7 eV below the Pr $M_5$ peak of the x-ray absorption as shown in Fig. 2**d**. Notably, the resonant energy profile spans a much broader energy range than that of Ni $L_3$ resonance. In addition, the resonant reflection shows up at the Pr $M_4$ absorption edge but not at the Ni $L_2$ absorption edge (see the Supplementary Materials for details).

After identifying the electronic resonance of the scattering peaks at ($\sim\frac{1}{3}$, 0, 0.35) in PrNiO$_2$, we have then investigated its intra-unit-cell symmetry. As photon energies are tuned at the Ni $L_3$ absorption edges ($2p_{x,y,z} \rightarrow 3d$), the resonant reflection becomes sensitive to charge modulation on the Ni sites and the bonding to neighboring oxygen ions because the charge modulation directly affects the final state of the $3d$ XAS process. As illustrated in Fig. 3**a**, a wedge-shaped sample holder was used to rotate the azimuthal angle $\phi$ around $\vec{Q}_{co}$. In Fig. 3**b**, we show real-space schematics of charge modulation symmetry components corresponding to a site-centered $s$-wave, an extended $s'$-wave, and a $d$-wave. For the $s$-wave, the extra charge sits in Ni $3d$ orbitals, while the $s'$- and $d$-waves are bond-type orders for which the spatially modulated charge density is on the O-$2p$ states. The symmetry components can be written as $\Delta_s$, $\Delta_{s'}(\cos k_x + \cos k_y)$, and $\Delta_d(\cos k_x - \cos k_y)$ for $s$-, $s'$-, and $d$-wave, respectively, as defined by theory in ref. [33]. Using a linear combination of these symmetry components (see the Supplementary Materials for details), we construct the scattering tensor $F_x$ to model the azimuthal angle dependence of RXS intensity according to $I_{RXS}(\vec{Q}, \omega) \propto \left| \sum_{pq} \epsilon_p \cdot \sum_x F_x(\omega) e^{-i\vec{Q}\cdot\vec{r}} \cdot \epsilon'_q \right|^2$ where $\epsilon$ ($\epsilon'$) represents the polarization vector for incoming (outgoing) photons, and the scattering tensor at the charge order wave vector is a diagonal tensor of the linear combination of the magnitudes of the symmetry components $\delta_s$, $\delta_{s'}$, and $\delta_d$, corresponding to $s$-, $s'$-, and $d$-wave, respectively (see the Supplementary Materials for further details) [35].

Figure 3**c** shows the experimental data of the azimuthal angle dependence of the background-subtracted RXS signals for the $\sigma$- and $\pi$-polarization of incident photons. The ratios of the integrated intensity (extracted from Gaussian fits to the data) $I_\pi/I_\sigma$ for different $\phi$ are shown in Fig. 3**d**. The self-absorption corrections are included in the calculated profiles to analyze the experimental data (see the Supplementary Materials for details). We first fit the data using a single symmetry component. The solid lines in the upper panel of Fig. 3**d** represent the best-fit results using the *s*-, *s'*-, and *d*-wave models, respectively. Yet, they all deviate substantially from the data. We then extend the model to all possible combinations of two symmetry components (*s* + *s'*, *s* + *d*, and *s'* + *d*). The best-fit results are given in the lower panel of Fig. 3**d** as the solid orange and blue curves, corresponding to the *s* + *d* and *s'* + *d* models, respectively. It shows significantly improved agreement with the data, while the fit using the *s* + *s'* model (the dashed line) still deviates considerably from the data. Although the present study can hardly distinguish between the *s* + *d* and *s'* + *d* models, the best-fit to the data yields $\delta_s/\delta_d = 0.09$ and $\delta_{s'}/\delta_d = 0.16$, indicating the predominant *d*-wave character of the charge order in both cases. We caution, however, that the proportion of *s*-, *s'*-, and *d*-wave symmetry depends on the energy and momentum of the electronic states; the azimuthal experiment with a lower detection angle yields enhanced sensitivity to the *s*-wave symmetry of the CDW order at the Cu *L* edge [38]. We therefore have calculated azimuthal angle dependence of RXS measurements for different detector angles, which shows our scattering geometry ($L = 0.35$, detection angle $2\theta = 150.2°$) provides significant contrast between *s*-, *s'*-, and *d*-form factor models (see the Supplementary Materials for details).

The predominant *d*-wave character of the superlattice reflection at the Ni $L_3$ absorption edge suggests that the charge modulation mainly resides on the O-2*p* orbitals, underscoring the crucial role of the O-2*p* ligand states in hole-doped infinite-layer nickelates. It seems, however, at odds with the fact that the parent phase of infinite-layer nickelates is a Mott-Hubbard insulator according to the Zaanen-Sawatzky-Allen (ZSA) scheme [26], for which extra charge would be expected to reside on Ni sites, resulting in an *s*-wave charge modulation. This seeming inconsistency may be reconciled by the fact that the Mott insulating $NiO_2$ layers lie in the critical crossover regime of the ZSA diagram where $S = 0$ and $S = 1$ eigenstates cross; in this critical regime, the extra charge in the $NiO_2$ layers may possess a strong O-2*p* component with the same $^1A_1$ symmetry like Zhang-Rice singlet (ZRS) of cuprates as proposed in theory [39, 40].

Despite the similarities in the local symmetry of charge order between infinite-layer nickelates and certain cuprates [34-36], some salient differences exist. First, the superlattice peaks in infinite-layer nickelates resonate at the rare-earth $M_{5,4}$ absorption edges [23], while in cuprates they appear not to be directly linked to the rare-earth elements. In addition, the integrated intensity of the resonant reflections diminishes gradually at the Ni $L_3$ resonance, while it remains unchanged at the Pr $M_5$ resonance from 20 K to 400 K, as shown in Fig. 4**b**; in both cases, the correlation length is about 40 Å and stays almost unchanged from 20 K to 400 K (see Fig. 4**c**). This is in contrast with that charge orders occur over a temperature range from 100 K to 150 K in underdoped cuprates [30]. Furthermore, the scattering at the Pr $M_5$ resonance peaks at $L = 0.365$ (r. l. u.) with a correlation length $\xi \sim 3.2$ Å along *L*, unlike in cuprates where the scattering of charge order is often maximized at half-integer in the unit of r. l. u. along *L*, indicating possible anti-phase modulations between neighboring layers [14]. Unfortunately, as shown in Fig. 4a, $L = 0.365$ (r. l. u.) coincides with the limit of reciprocal space accessible at the Ni $L_3$ resonance, and therefore the out-of-plane modulation of the superlattice reflection at the Ni $L_3$ resonance remains undermined.

The resonant reflections at the rare-earth sites in $La_{1-x}Sr_xNiO_2$ were proposed as a secondary effect inherited from the charge modulation in $NiO_2$ planes [23], which, however, cannot explain the apparent dissimilarities from those at the Ni $L_3$ resonance. Alternatively, the reflections at the rare-earth $M_5$ resonance may have fundamentally different origins from those at the Ni $L_3$ resonance. A possible scenario to explain these dissimilarities is that small amounts of residual oxygen after the topotactic reduction process can contribute partly to the charge modulations, given that the resonant reflection is observed in an incompletely reduced film (see the Supplementary Material for further information), which has a relatively larger $c$ lattice constant likely owing to the excess oxygen [41]. The nominally stoichiometric 'parent' phase may be better represented by $PrNiO_{2+\delta}$, where $\delta$ is likely small. The residual oxygen may organize into an ordering pattern and simultaneously dope holes into the $NiO_2$ planes, which explains the observation of superconductivity in $LaNiO_2$ [42] and charge orders in $RNiO_2$ (R= La and Nd) [23-25]. This somewhat resembles the excess oxygen dopants in $La_2CuO_{4+\delta}$, in which they occupy interstitial sites forming an ordering pattern while simultaneously doping holes into the $CuO_2$ planes [43, 44]. In infinite-layer nickelates, a small amount of apical oxygen in the $NiO_6$ octahedra may not be removed and arrange themselves in an ordered state as revealed at the Pr $M_5$ absorption edge. However, the ordering of excess oxygen itself cannot explain the discrepancy between the resonant reflection measured at the Ni $L_3$ and that at the Pr $M_5$ absorption edges, especially in the temperature dependence and the local symmetry. Alternatively, the residual oxygen can dope holes into the $NiO_2$ planes and that may account for the charge order disclosed at the Ni $L_3$ absorption edge.

The above scenario recalls the identification of distinct charge orders in the $CuO_2$ planes and chains of $YBa_2Cu_3O_{6+\delta}$ [15]. The enhancement of the resonant reflection for the oxygen ordering in the chains gets maximized at ~2.4 eV above the $L_3$ peak of the x-ray absorption, while for the charge density wave in the $CuO_2$ planes, the resonance peaks ~0.1 eV below the $L_3$ peak. By contrast, the resonant reflections in $PrNiO_{2+\delta}$ become strongest respectively at ~0.7 eV below the Ni $L_3$ peak and ~0.7 eV below the Pr $M_5$ peak of the x-ray absorption. This same amount of shift toward lower energy indicates the charge modulations at the two resonances are probably intrinsically linked. The energy shift referring to the peak A′ is ~0.3 eV at the Ni $L_3$ absorption edge, which is comparable to that of ~0.1 eV shift for the charge density wave in the $CuO_2$ planes [15], but much smaller than ~2.4 eV shift for the oxygen ordering in the chains. In the case of $La_{1-x}Sr_xNiO_2$, the resonant reflection is ~0.3-0.5 eV below the Ni $L_3$ peak, while its shift relative to the La $M_5$ peak seems hard to resolve owing to the overlap between the La $M_5$ and the Ni $L_3$ peaks of x-ray absorption [23]. The resonance profile associated with the superlattice reflection in $PrNiO_{2+\delta}$ agrees well with that in $LaNiO_2$ at the Ni $L_3$ absorption edge.

In summary, we have systematically investigated the resonant reflections at the Ni $L_3$ and Pr $M_5$ absorption edges, revealing significant dissimilarities in the dependence on temperature, photon energy, $L$, and azimuthal angle between the two resonances. We propose that excess oxygen should account for the emergence of resonant reflections revealed at the Ni $L_3$ and Pr $M_5$ absorption edges. This is further supported by recent STEM measurements on $NdNiO_2$ films that show oxygen intercalations with a similar wave vector as that revealed by resonant x-ray scattering [45, 46]. In particular, the azimuthal angle dependence analysis of the resonant reflection at the Ni $L_3$ absorption edge underscores the crucial role of oxygen ligands in the charge modulation at the $NiO_2$ planes, remindful of that in certain cuprates [33-35]; by contrast, the charge modulation at the Pr $M_5$ absorption edge is compatible with that the extra charge primarily occupies the Pr sites via bonding to excess oxygen dopants. The scattered intensity of the resonant reflection at the Ni $L_3$ resonance decreases very gradually as the temperature increases, a 40% decrease from 20 K to 400 K, and remains finite at 400 K, while at the Pr $M_5$ resonance, the scattering intensity does not vary across the whole temperature range surveyed in this study. This agrees well with the early study on the $LaNiO_{2+\delta}$. The fact that the charge modulation at the Ni $L_3$ resonance develops at such a high temperature may pertain to the Pr-Ni hybridization. The different dopant ions, i.e., excess oxygen versus $Sr^{2+}$, could pertain to the rapid decrease of the onset temperature of the charge order in $La_{1-x}Sr_xNiO_2$ compared to $LaNiO_{2+\delta}$ [23]. This is in line with the theoretical study showing that hole doping with Sr in $NdNiO_2$ tends to make the material more cuprate-like as it minimizes the self-doping effect [47]. At odds with the previous report that the charge order was observed only in an uncapped $NdNiO_2$ film [24], which raises questions about the role of the surface or interface state due to the capping layer, our observation of the charge order in a capped $PrNiO_{2+\delta}$ film suggests that the capping might not be critical for the emergence of charge order in infinite-layer nickelates.

**METHODS**

Thin film samples of (0 0 1)-oriented $PrNiO_3$ with $SrTiO_3$ capping layer have been synthesized on (0 0 1) $SrTiO_3$ substrates via pulsed laser deposition (PLD) using a KrF excimer laser ($\lambda = 248$ nm). The infinite-layer $PrNiO_2$ was obtained by a soft-chemistry reduction process using $CaH_2$ powder [5, 48, 49]. The $SrTiO_3$ films were deposited on the $PrNiO_3$ films as the capping layers immediately after the deposition of the precursor films under the same condition. The $PrNiO_3$ films capped with the $SrTiO_3$ top layers were then used for the topotactic reduction process to obtain an infinite-layer phase. The RXS experiments were performed at the Resonant Elastic Inelastic X-ray Scattering (10-ID2) of the Canadian Light Source [50], equipped with a 4-circle diffractometer in a $10^{-10}$ mbar ultrahigh-vacuum chamber. The photon flux is about $5\times10^{11}$ photons per second and energy resolution reaches $\Delta E/E \sim 2\times10^{-4}$. The incoming photons are fully polarized with two configurations $\sigma$ and $\pi$, corresponding respectively to the polarization vector perpendicular to and in the scattering plane.

**DATA AVAILABILITY**

The data will be released upon final publication.

## ACKNOWLEDGEMENT

We thank Guang-Ming Zhang and Fu-Chun Zhang for fruitful discussions. This work was supported in part by the National Natural Science Foundation of China (Grant No. 12074411), the National Key Research and Development Program of China (Grant No. 2022YFA1403900, 2021YFA1401800 and 2017YFA0302900), the Strategic Priority Research Program (B) of the Chinese Academy of Sciences (Grant No. XDB25000000), the Swiss National Science Foundation (Grant 200021_188564) and the Research Program of Beijing Academy of Quantum Information Sciences (Grant No. Y18G06). Part of the research described in this paper was performed at the Canadian Light Source, a national research facility of the University of Saskatchewan, which is supported by the Canada Foundation for Innovation (CFI), the Natural Sciences and Engineering Research Council (NSERC), the National Research Council (NRC), the Canadian Institutes of Health Research (CIHR), the Government of Saskatchewan, and the University of Saskatchewan. This work used resources of the National Synchrotron Light Source, a U.S. Department of Energy (DOE) Office of Science User Facility operated for the DOE Office of Science by Brookhaven National Laboratory under Contract No. DE-AC02-98CH10886 and the Laboratory Directed Research and Development project of Brookhaven National Laboratory under Contract No. 21-037.

## AUTHOR CONTRIBUTIONS

Z.H.Z. and X.J.Z conceived the research. X.L.R. and Q.G. prepared and characterized the film samples. X.L.R., R.S., Q.G., and Z.H.Z. performed the RXS experiments. X.L.R and Z.H.Z. analyzed the data. Z.H.Z., X.J.Z., T.X., J.H., J.C., R.C., Y.W., Q.W., J.L., and X.L.R.discussed and interpreted the results. X.L.R. and Z.H.Z. wrote the manuscript with input from all authors.

**DECLARE OF INTEREST**

The authors declare no competing interests.

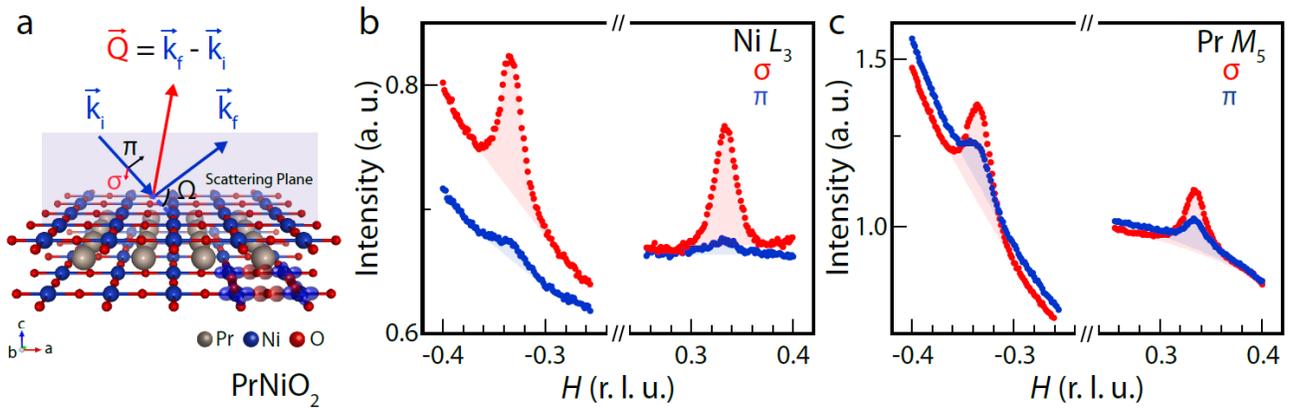

FIG. 1: Schematic of the scattering geometry and resonant enhancement of $(\pm\frac{1}{3}, 0, 0.365)$ reflection at the Ni $L_3$ and the Pr $M_5$ absorption edges. **a**, Crystal structure and RXS measurement geometry. **b** and **c** show momentum scans across wavevector $\vec{Q}_{CO} = (\pm\frac{1}{3}, 0, 0.365)$ using a photon energy of 851.3 eV and 928.4 eV, respectively, for a typical PrNiO$_2$ film; asymmetry comes from the fluorescence background in the specific scattering geometry [14-16, 24, 25].

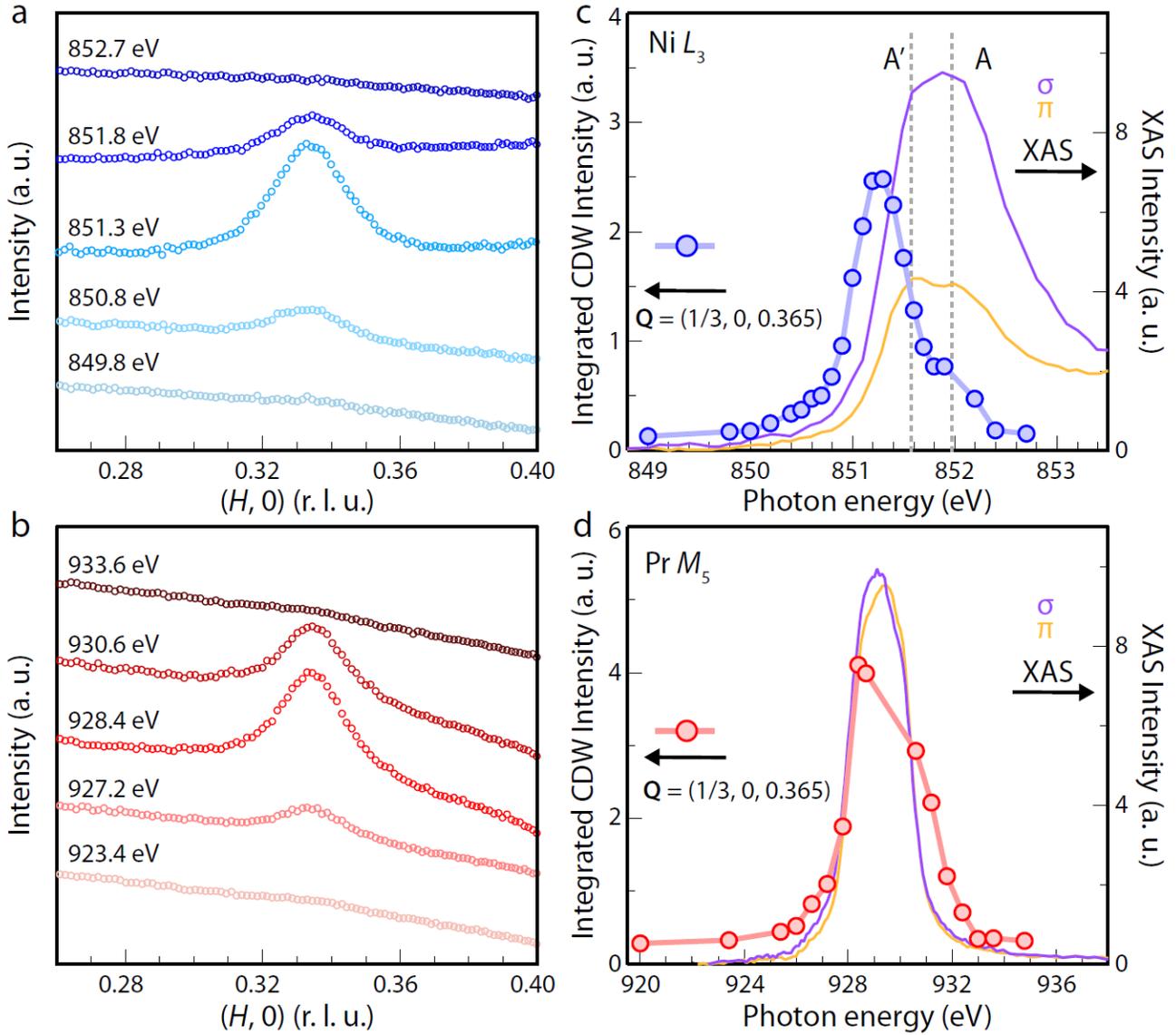

**FIG. 2: Resonant enhancement of the charge order reflection at the Ni $L_3$ and Pr $M_5$ absorption edges. a** and **b** show a series of representative momentum scans across wavevector ($\frac{1}{3}$, 0, 0.365) for different photon energies with $\sigma$ polarization of incoming photons, where both $\theta$ and $2\theta$ angles were adjusted for each different photon energy such that the center of the scans corresponds to the constant wave vector ($\frac{1}{3}$, 0, 0.365). **c** and **d** represent the integrated intensity of the resonant reflection (extracted by using Gaussian fits to the momentum scans) as a function of the photon energies near the Ni $L_3$ and the Pr $M_5$ absorption edges, respectively, together with the XAS collected using different photon polarizations (right scale). All the measurements were taken at 20 K.

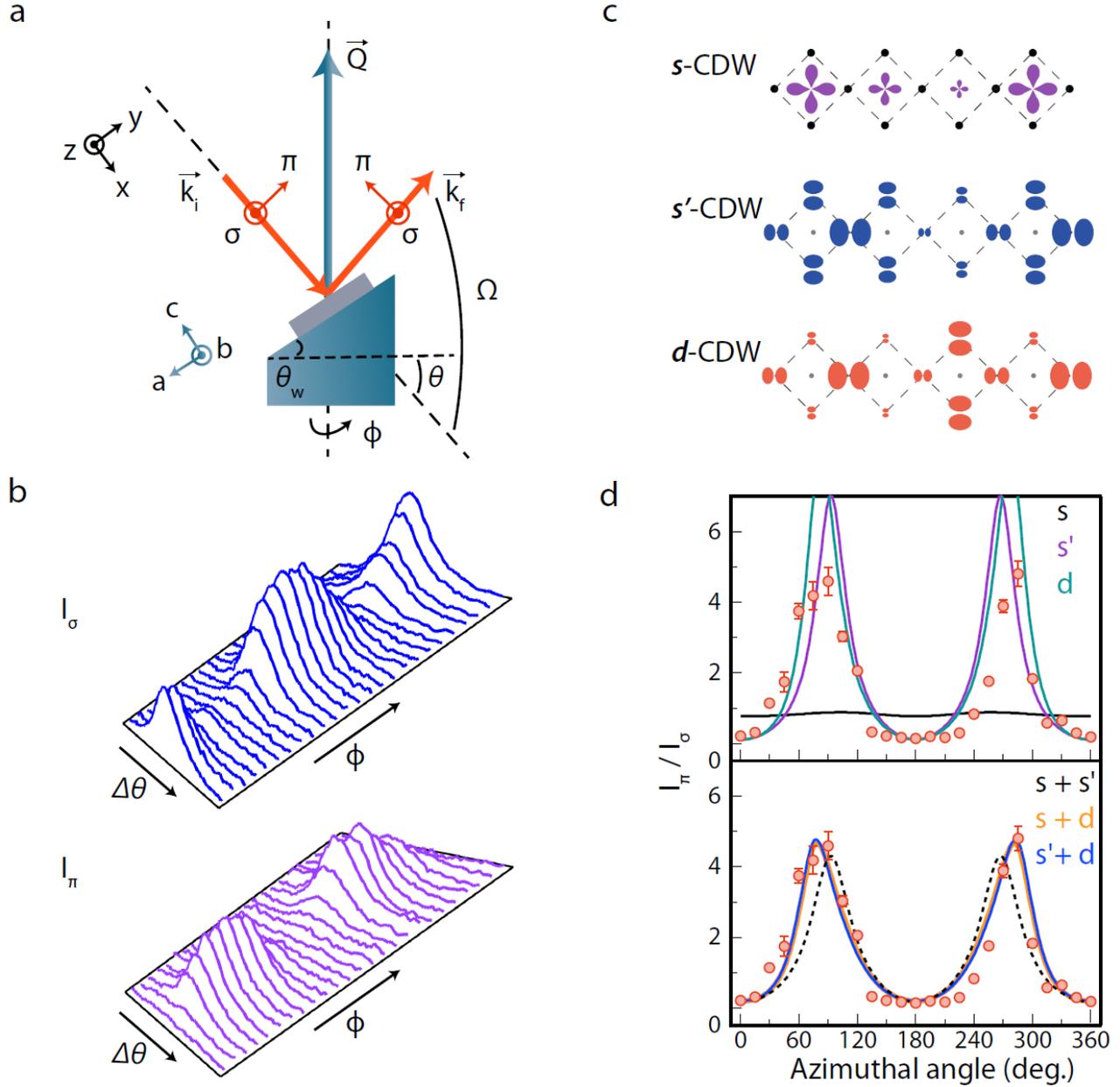

**FIG. 3: Scattering geometry, schematics of charge modulation symmetry, and azimuthal angle dependence of RXS measurements. a**, Top view of the experimental geometry in the laboratory frame. $\vec{k}_i$ and $\vec{k}_f$ represent the incoming and outgoing photon wavevectors; the transferred momentum $\vec{Q}$ is defined as $\vec{k}_f - \vec{k}_i$. The scattering angles are defined by the sample rotation $\theta$, the detector angle $\Omega$, the azimuthal rotation angle $\phi$, and the wedge angle $\theta_w$. **b**, Azimuthal angle dependence of the momentum scans of the CDW peak at $\vec{Q} \sim (\frac{1}{3}, 0, 0.35)$ in PrNiO$_2$ at the photon energy of 851.3 eV. **c**, Real-space schematics of charge modulation symmetry components: $s$-, $s'$-, and $d$-wave local symmetry. **d**, The scattering intensity ratio $I_\pi/I_\sigma$ from the measurements is fitted by using a single symmetry component (top panel) and a mixture of two symmetry components (lower panel), respectively.

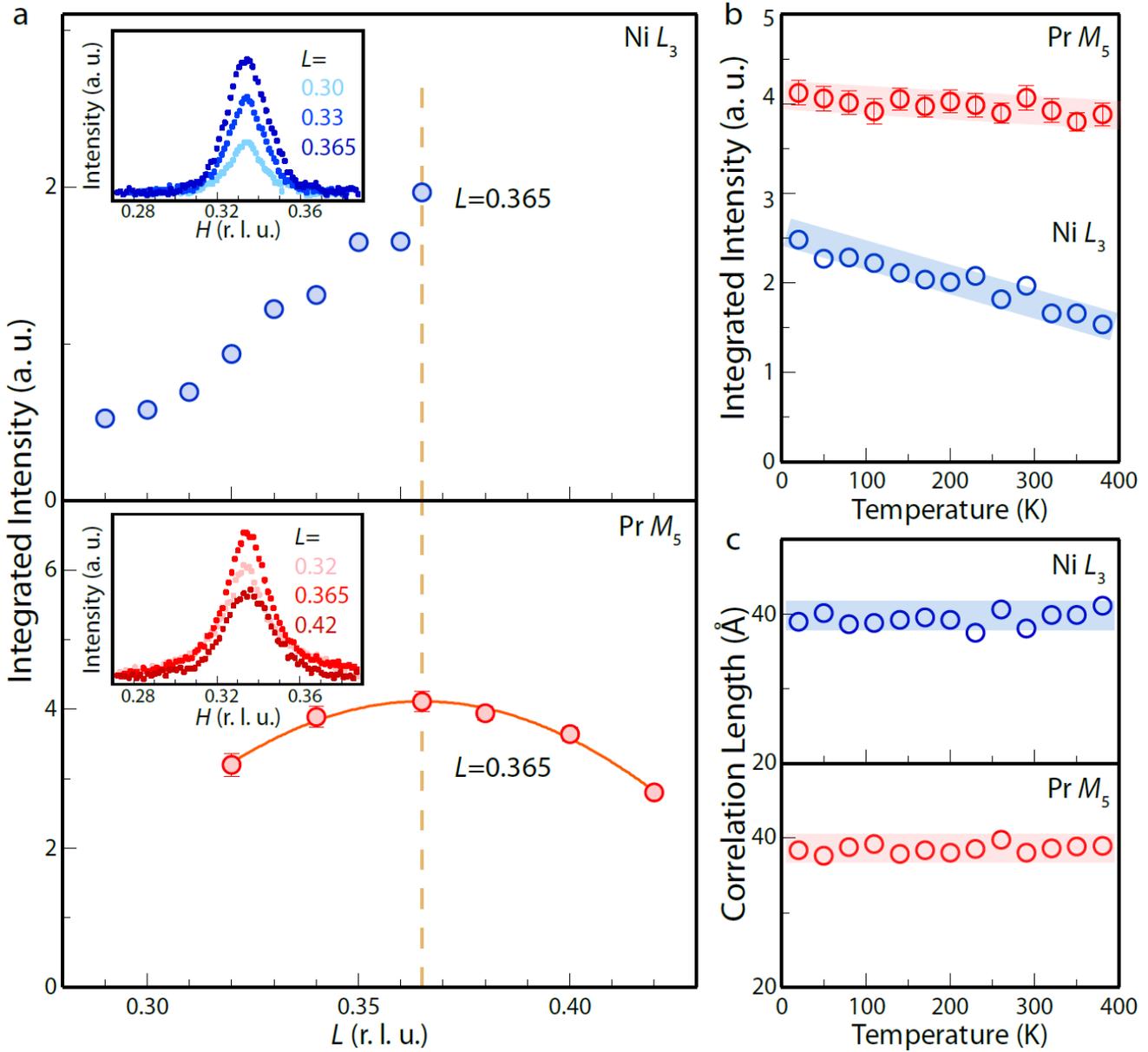

**FIG. 4: *L* and temperature dependence of the CDW order at the Ni *L*$_3$ and the Pr *M*$_5$ absorption edges.** **a**, shows *L* dependence of the scattering intensity for the resonant reflection of the CDW order at 851.3 eV (the upper panel) and 928.4 eV (the lower panel), respectively; the insets are representative momentum scans for different *L*. **b** and **c** summarize the temperature dependence of the scattered intensity and the correlation length for the CDW order, respectively.